\documentclass[sigconf]{acmart}

\usepackage{booktabs} % For formal tables

\setcopyright{acmlicensed}
\acmConference[LND4IR '18]{ACM SIGIR 2018 Workshop on Learning from Limited or Noisy Data}{July 12, 2018}{Ann Arbor, Michigan, USA} 
\acmYear{2018}
\copyrightyear{2018}
\acmDOI{}
\acmISBN{}
\acmPrice{}

\begin{document}
\title{Information Retrieval in African Languages}
%\titlenote{Produces the permission block, and
%  copyright information}
\subtitle{Position Paper}
%\subtitlenote{The full version of the author's guide is available as
%  \texttt{acmart.pdf} document}

\author{Hussein Suleman}
%\authornote{Dr.~Trovato insisted his name be first.}
\orcid{0000-0002-4196-1444}
\affiliation{%
  \institution{University of Cape Town}
  \streetaddress{18 University Avenue}
  \city{Cape Town}
  \country{South Africa}
  \postcode{7701}
}
\email{hussein@cs.uct.ac.za}

\renewcommand{\shortauthors}{H. Suleman}

\begin{abstract}
Developing Information Retrieval (IR) tools and techniques in African 
languages suffers from the dual problems of a lack of algorithms
and very small test data collections.  This affects the 
creation of practical IR systems and limits the ability to
apply IR to address human and socio-economic problems, which is an urgent
need in poor countries.  This
position paper presents an overview of recent and current
work conducted at the University of Cape Town in this area.
While many problems have been investigated at an early stage, 
limited dataset sizes for local African languages still
persists as a significant limitation and stumbling block.
\end{abstract}

\begin{CCSXML}



\maketitle

\section{Introduction}

In countries where English is not the only
language spoken, algorithmic support for non-English content varies. In poor
African countries, very little is known about
searching in local languages.  This impacts on the use of local languages
for teaching and learning, knowledge discovery and, especially, the
addressing of development-centric problems in such countries.  Recent
research done at the University of Cape Town has centred on the duality of
exploring Information Retrieval (IR) in African languages, with a focus on Bantu
languages, and the use of IR to explicitly address human and socio-economic
development problems in poor countries.

\section{Recent Research}

Early explorations in multilingual document collections made it apparent that there was
a miniscule online presence in languages such as isiZulu and isiXhosa (the largest South African language
groups, with 8-9 million speakers of each)
and creating new document collections was non-trivial \cite{Poulo09}.  Using
Wikipedia as a gross estimate for interest in electronic document creation,
most South African languages contain approximately 1000 documents or fewer,
as of 2018, and appear in
the lowest quartile of languages by content.

Our first major study related to African multilingualism discovered that most commercial search engines make a single language assumption about queries
so users who are fluent in multiple languages do not get good results from mixed language queries
\cite{Mustafa11}.  This study went on to show how a deeper understanding of queries and documents without single-language
assumptions can support a better quality re-ranking of documents. Arabic was the language of this study because of
the scarcity of documents in isiXhosa or isiZulu around 2007, and limited interest in corpus development and 
computational linguistics in Bantu languages at the time.
 
Being invited to participate in the EU MUMIA Cost Action (2012-2014) made it apparent that low resource languages 
are, in fact, a common problem internationally. In addition, given renewed interest in
local language issues \cite{Pretorius09} \cite{Keet14}, search in African languages seemed more viable.

A case study was then
conducted into the feasibility of an isiZulu search engine \cite{Malumba15} and it was clear that language preprocessing
tools and language identification tools were the key elements needed for basic non-English support. 
The focus was on algorithms such as stemmers and language detection that
used a combination of statistical (e.g., n-gram) and natural language processing
(e.g., morphological analysis) techniques.
Follow-on projects have considered baseline search engines for isiXhosa
\cite{Kyeyune15} and, more recently, ciShona. 

In parallel with developing IR tools for low-resource Bantu languages, we
have also considered how users will access these tools in poor countries. 
Assuming that such users only have access to mobile devices, the feasibility
of a fully-isiXhosa speech-driven interface on a mobile phone was
demonstrated \cite{Modise17}.

\section{Ongoing and Future Work}

Because of natural similarities between these languages, and the high number of
regional languages that are considered low resource languages, a broader project is attempting to build language group-
oriented tools (such as stemmers \cite{Chavula17}) to exploit similarity of the languages at the
processing stages \cite{Chavula16} and exploit the fact that users who can read one language can often
read many related languages as well. 

There are ongoing efforts to collect original and translated
texts in multiple low-resource local languages, 
both as an end in itself and to support further research in this area
\cite{Vonholy17}. We are using multiple variations of gamification techniques 
to develop these 
corpora for low-resource languages \cite{Packham15}, where is seems that
gamification for corpus development works somewhat differently in poor
communities than has been reported elsewhere.

Finally, assuming that limited test data will always be a constraint,
we are considering how language identification in
low-resource Bantu languages works as a function of language model scarcity 
and unidentified text sparsity.

A parallel strand of research is considering information-centric solutions 
to address human and
socio-economic problems.  Early work is looking at how users find jobs and
how levels of development may be monitored computationally using 
combinations of IR and data mining approaches \cite{Mwanza16} \cite{Mwanza17}.

Ultimately, the goal is to develop African language IR in parallel with
other information-centric efforts to actively address development-oriented 
problems in poor countries.

\begin{acks}
This research was partially funded by the National Research Foundation
of South Africa (Grant numbers: 85470 and 88209) and University
of Cape Town. The authors acknowledge that opinions,
findings and conclusions or recommendations expressed in this
publication are that of the authors, and that the NRF accepts no
liability whatsoever in this regard.
\end{acks}

\bibliographystyle{ACM-Reference-Format}

\bibliography{lnd4ir}

\end{document}